\documentclass[10pt,preprint2]{aastex}
\shorttitle{Measuring Oblateness for Transiting EGPs}
\shortauthors{Barnes \& Fortney}

\begin{document}

\title{Measuring the Oblateness and Rotation of Transiting Extrasolar
Giant Planets}

\author{Jason W. Barnes and Jonathan J. Fortney}
\affil{Department of Planetary Sciences, University of Arizona, Tucson, AZ, 85721}
\email{jbarnes@c3po.lpl.arizona.edu, jfortney@lpl.arizona.edu}

\newpage

\begin{abstract} 

We investigate the prospects for characterizing extrasolar giant planets by
measuring planetary oblateness from transit photometry and inferring planetary
rotational periods.  The rotation rates of planets in the solar system vary
widely, reflecting the planets' diverse formational and evolutionary histories. 
A measured oblateness, assumed composition, and equation of state yields a
rotation rate from the Darwin-Radau relation.  The lightcurve of a transiting
oblate planet should differ significantly from that of a spherical one with the
same cross-sectional area under identical stellar and orbital conditions. 
However, if the stellar and orbital parameters are not known \emph{a priori},
fitting for them allows changes in the stellar radius, planetary radius, impact
parameter, and stellar limb darkening parameters to mimic the transit signature
of an oblate planet, diminishing the oblateness signature.  Thus even if
HD209458b had an oblateness of $0.1$ instead of our predicted $0.003$, it would
introduce a detectable departure from a model spherical lightcurve at the level
of only one part in $10^5$.  Planets with nonzero obliquity break
this degeneracy because their ingress lightcurve is asymmetric relative to that
from egress, and their best-case detectability is of order $10^{-4}$. 
However, the measured rotation rate for these objects is non-unique due to
degeneracy between obliquity and oblateness and the unknown component of
obliquity along the line of sight.  Detectability of oblateness is maximized for
planets transiting near an impact parameter of $0.7$ regardless of obliquity. 
Future measurements of oblateness will be challenging because the signal is near
the photometric limits of current hardware and inherent stellar noise levels.

\end{abstract}


\section{INTRODUCTION}

Discovery of a transiting extrasolar planet, HD209458b
\citep{2000ApJ...529L..45C, 2000ApJ...529L..41H}, has provided one mechanism for
researchers to move beyond the discovery and into the characterization of
extrasolar planets.  Precise \emph{Hubble Space Telescope} measurements of
HD209458b's transit lightcurve revealed the radius ($1.347 \pm 0.060
\mathrm{R_{Jup}}$) and orbital inclination ($86.68^\circ \pm 0.14^\circ$, from
impact parameter $0.508$) of the planet, which, along with radial velocity
measurements, unambiguously determine the planet's mass ($0.69 \pm 0.05
\mathrm{M_{Jup}}$) and density 
\citep[$0.35 \mathrm{g~cm^{-3}}$; ][]{2001ApJ...552..699B}.  Of the over 100
extrasolar planets discovered so far, this large radius makes
HD209458b the only one empirically determined to be a gas giant.  

Knowledge of a planet's cross-sectional area provides a zeroeth order
determination of its geometry, and the \emph{HST} photometry is precise enough
to constrain the shape of HD209458b to be rounded to first order.  While a
planet transits the limb of its star, the rate of decrease in apparent
stellar brightness is related to the rate of increase in  stellar surface area
covered by the planet in the same time interval.  We investigate whether it is
possible to use this information to determine the shape of the planet to second
order.

Rotation causes a planet's shape to be flattened, or oblate, by reducing the
effective gravitational acceleration at the equator (as a result of centrifugal
acceleration) and by redistributing mass within the planet (which changes the
gravity field).  Oblateness, $f$, is
defined as a function of the equatorial radius ($R_{eq}$) and the polar radius
($R_p$):
\begin{equation}
\label{eq:f}
f \equiv \frac{R_{eq}-R_{p}}{R_{eq}} \mathrm{.}
\end{equation}
For Jupiter and Saturn, high rotation rates and low densities result in highly
oblate planets: $f_{Jupiter} = 0.06487$ and $f_{Saturn} = 0.09796$, compared to
$f_{Earth} = 0.00335$.  

An earlier investigation of the detectability of oblateness in transiting
extrasolar planets was published by \citet{Seager.oblateness}.  Our work
represents an improvement over \citet{Seager.oblateness} in the use of model
fits to compare oblate and spherical planet transits, the use of the
Darwin-Radau relation to associate oblateness and rotation, and a thorough
investigation of the degeneracies involved in fitting a transit lightcurve.

In this paper, we investigate the reasons for measuring planetary rotation
rates, the relationship between rotation rate and oblateness for extrasolar
giant planets, the effect of oblateness on transit lightcurves, and the
prospects for determinating the oblateness of a planet from transit photometry.

\begin{figure} 
\plotone{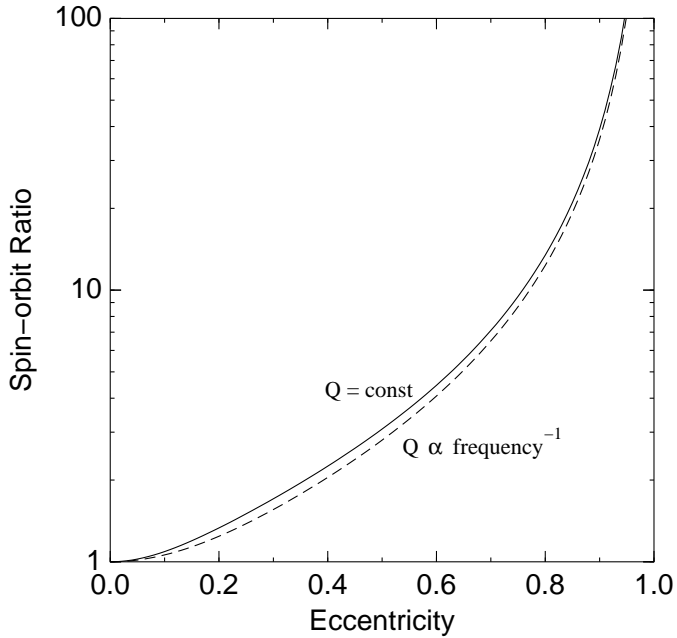}
\caption{Spin to orbital mean motion ratios for tidally evolved fluid planets in
equilibrium.  The solid line represents the ratio as calculated under the
assumption of a frequency-indenpendant tidal dissipation factor $Q$, and the
dashed line is calculated assuming $Q \propto \mathrm{frequency}^{-1}$.  Under
these assumptions, extremely eccentric planet HD80606b ($e=0.93$) would, if
allowed to come to tidal equilibrium, rotate over $90$ times faster than its
mean orbital motion! \label{figure:spinorbit}} 
\end{figure} 

\begin{figure} 
\plotone{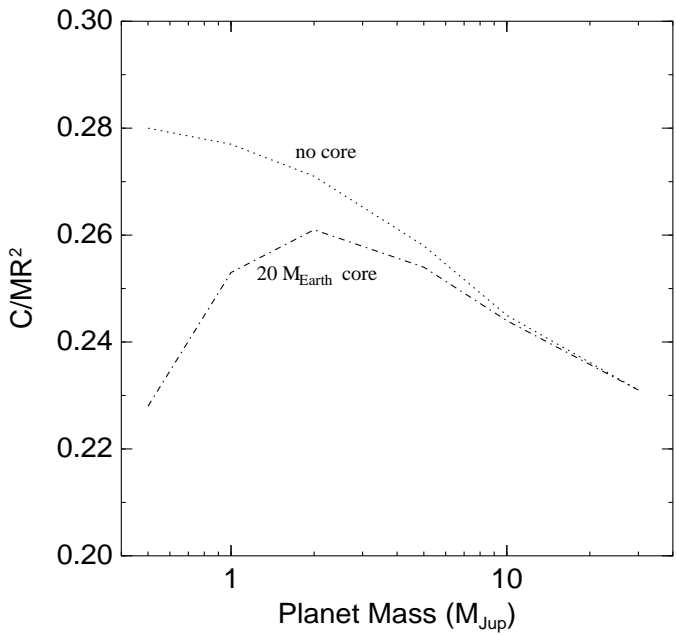}
\epsscale{0.5}
\caption{The moment of inertia coefficient, $\mathbb{C}$, as a function of
planet mass for hypothetical generic $1.0~\mathrm{R_{Jup}}$ extrasolar giant
planets.  Extrasolar giant planets may or may not possess rocky cores
depending on their formation mechanism, so we plot $\mathbb{C}$ for both no core
(upper curve) and an assumed $20~\mathrm{M_\oplus}$ core (lower curve). 
\label{figure:C_MR2}} 
\end{figure}

\begin{figure} 
\plotone{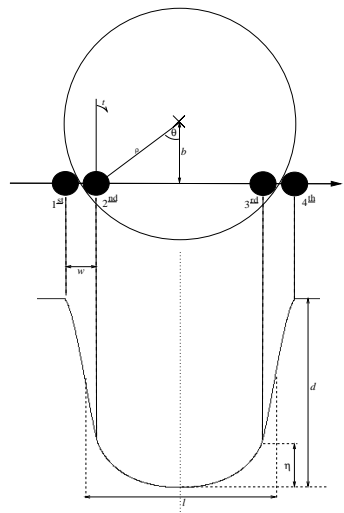}
\epsscale{0.9}
\caption{Anatomy of a transit, after BCGNB.\label{figure:schematic}} 
\end{figure}

\begin{figure} 
\plotone{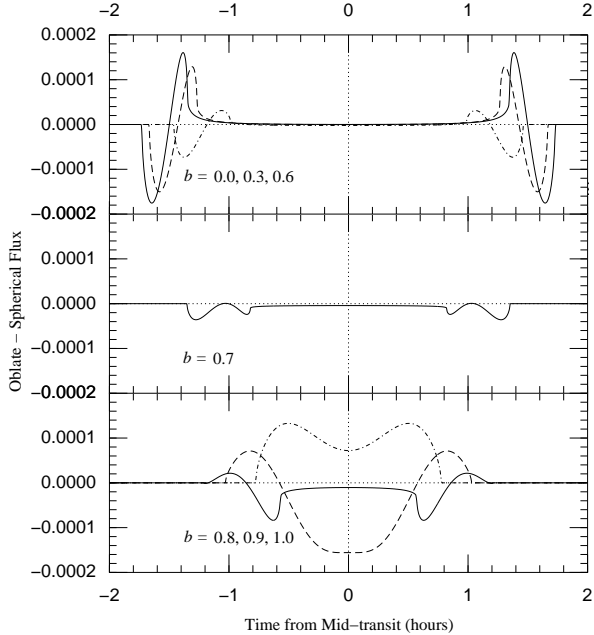}
\epsscale{0.9}
\caption{Oversimplified model of the effect of oblateness on a transit
lightcurve.  We plot the differences between the lightcurve of an oblate planet
and the lightcurve of a spherical planet with the same cross-sectional area,
$F_{f=0.1}(t) - F_{f=0.0}(t)$ while holding all other transit parameters, $R_*$,
$R_p$, $b$, and $c_1$ the same and setting these values equal to those values
measured by BCGNB for HD209458b.  The top panel plots the lightcurve
differential for impact parameters $b=0.0$ (solid line), $b=0.3$ (dashed line),
and $b=0.6$ (dot-dashed line).  An oblate planet for $b<0.7$ encounters first
contact before, and second contact after, the equivalent spherical planet,
resulting in the initial negative turn for $F_{f=0.1}(t) - F_{f=0.0}(t)$, and
subsequent positive section of the curve for these impact parameters.  In the
bottom panel, we plot the differential lightcurve for $b=0.8$ (solid line),
$b=0.9$ (dashed line), and $b=1.0$ (dot-dashed line).  For $b>0.7$, under
otherwise identical conditions, an oblate planet first encounters the limb of
the star after the equivalent spherical planet, and last touches the limb later
on ingress.  For $b=0.9$, because $R_p>0.1~R_*$ there is no second contact, i.e.
the transit is partial, so the flux differential does not return to near zero,
even at mid-transit.  The middle plot shows the
lightcurve differential at the changeover point between these two regimes, where
$b=0.7$.
\label{figure:transit_effects}}
\end{figure}

\section{EXOPLANETS AND ROTATION}

While knowledge of a planet's mass and radius provides information regarding
composition and thermal evolution, measurements of rotation and obliquity 
promise to constrain the planet's formation, tidal evolution, and tidal
dissipation.  What these data might reveal about a planet depends
on whether the planet is unaffected by stellar tides, slightly affected by
tides, or heavily influenced by tides.

\subsection{Tidally Unaffected}

The present-day rotation rate of a planet is the product of both the planet's
formation and its subsequent evolution.  Planets at sufficiently large distances
from their parent stars are not significantly affected by stellar tides, thus
these objects rotate with their primordial rates altered by planetary
contraction and gravitational interactions between the planet, its satellites,
and other planets.  To the degree that a planet's rotational angular momentum is
primordial, it may be diagnostic of the planet's formation.  Planets formed in
circular orbits from a protoplanetary disk inherit net prograde angular momentum
from the accreting gas, resulting in rapid prograde rotation
\citep{1995Icar..114..217L}.  Planets that form in eccentric orbits receive less
prograde specific angular momentum than planets in circular orbits, and as a
result they rotate at rates varying from slow retrograde up to prograde
rotations similar to those of circularly accreted planets
\citep{Lissauer.1997.RotationfromEccentricAccretion}.  Thus, comparing the
current-day rotation rates for planets in circular and eccentric orbits may
reveal whether extrasolar planets formed in eccentric orbits or acquired their
eccentricity later from dynamical interactions with other planets or a disk.

The orientation of a planet's rotational axis relative to the vector
perpendicular to the orbital plane, the planet's obliquity or axial tilt $t$,
can also provide insight into the planet's formation mechanism.  Jupiter's low
obliquity ($3.12^\circ$) has been suggested as evidence that its formation was
dominated by orderly gas flow rather than the stochastic impacts of accreting
planetesimals \citep{1993ARA&A..31..129L}.  Tidally unevolved extrasolar planets
determined to have high obliquities could be inferred either to have formed
differently than Jupiter or to have undergone large obliquity changes as has
been suggested for Saturn \citep[$t=26.73^\circ$;][]{2002WardDPS}.

\subsection{Tidally Influenced}
 
Tidal interaction between planets and their parent stars slows the rotation of
those planets with close-in orbits \citep{1996ApJ...459L..35G}.  This tidal
braking continues until the net tidal torque on the planet becomes zero. 
Whether a planet reaches this end state depends on its age, radius, semimajor
axis, and the planet : star mass ratio.  

The rate of tidal braking also depends on the parameter $Q$, which represents
the internal tidal dissipation within the planet.   The value of $Q$ is poorly
constrained even for the planets in our own solar system
\citep{1966Icar....5..375G}.  Nevertheless, measurements of extrasolar planet
rotation rates could constrain $Q$ for these planets based on the degree of
tidal evolution that has taken place \citep{Seager.oblateness}.  

Tidal braking for objects with nonzero obliquity can, somewhat
counterintuitively, act to increase an object's obliquity.  Tidal torques
reduce the component of a planet's angular momentum perpendicular to the
orbital plane faster than they reduce the component of the planet's angular
momentum in the orbital plane \citep{Peale1999.ARAA.Origin&EvolutionofSats}. 
This occurs because at solstice the planet's induced tidal bulge is not carried
away from the planet's orbital plane by planetary rotation.  Therefore for
large fractions of the year stellar tidal torques do not act to right the
planet's spin axis, while torques that reduce the angular momentum
perpendicular to the orbital plane act year-round.  As a result, planets that
have undergone partial tidal evolution can exhibit temporarily increased
obliquity as the planet's rotation rate decreases.  Eventually, such a planet
reaches maximum obliquity and thereafter approaches synchronous rotation and
zero obliquity simnultaneously.  Planets undergoing tidal evolution may be
expected to have higher obliquities on average than planets retaining their
primordial obliquity.

\subsection{Tidally Dominated} \label{section:TidallyDominated}

The end state of tidal evolution for planets in circular orbits is a $1:1$
spin-orbit synchronization between the planet's rotation and its orbital
period, along with zero obliquity.   However, most of the extrasolar planets
discovered thus far are on eccentric orbits \citep{2000prpl.conf.1285M} (we
sometimes shorten 'planets on eccentric orbits' to 'eccentric planets' even
though the eccentricity is not inherent to the planet).  Thus these eccentric
planets will never reach $1:1$ spin-orbit coupling as a result of tidal
evolution because the tidal torque (Eq. \ref{eq:tidaltorque}) on the planet
from its star is much stronger (due to the $r^{-6}$ dependence) near
periapsis.  

The tidal torque between a planet and star is given by \citep{SolarSystemDynamics}
\begin{equation} \label{eq:tidaltorque}
\tau_{p-*}~=~-\frac{3}{2} \frac{k_{2}~G~M_*^2~R^5}{Q~r^6}
~\mathrm{sgn}(\Omega-\dot{\phi}) ~\mathrm{,}
\end{equation}
where $k_{2}$ is the planet's Love number, $R$
its radius, $\Omega$ its rotation rate (in radians per second), $\dot{\phi}$ the
instantaneous orbital angular velocity (also in radians per second), and $r$ is
the instantaneous distance between the planet and star.  The function
$\mathrm{sgn}(x)$ is equal to $1$ if $x$ is positive and $-1$ if $x$ is
negative.  The magnitude of the stellar perturbation is proportional to $GM_*^2$,
the product of the stellar mass ($M_*$) squared and the gravitational constant ($G$). 
The planet is spun down by tidal torques if its rotation is faster than its orbital
motion ($\Omega > \dot{\phi}$), and it is spun up if its rotation is slower than
the orbital motion ($\Omega < \dot{\phi}$).

If an eccentric planet were in a $1:1$ spin-orbit state, it would be spun up by
the star when its orbital angular velocity is greater than average near
periapsis, and it would be spun down when the orbital angular velocity is low
near apoapsis.  The total positive tidal torque induced while the planet is in
close will exceed the negative torque while it is far away, despite the shorter
time spent near periapsis.  Thus to be in rotational equilibrium with respect to
stellar tides, an eccentric, fluid planet must rotate faster than its mean
motion.

The Earth's moon avoids this fate because it has a nonzero component of its
quadrupole moment in its orbital plane.  The torque that the Earth exerts on
this permanent bulge exceeds the net tidal torque imparted on the moon due to
its eccentric orbit, keeping the Moon in synchronous rotation
\citep{SolarSystemDynamics}.  Fluid planets, however, have no permanent
quadrupole moment and thus have no restoring torque competing with the stellar
tidal torques \citep{1984Icar...58..186G}.  Tidal evolution ceases for these
bodies when the net torque per orbit is zero, which can only be achieved by
supersynchronous rotation.

The precise rotation rate necessary to balance the tidal torques over the
eccentric orbit depends on how $Q$ varies with the tidal forcing frequency (the
difference between the rotation rate and instantaneous orbital angular velocity,
$\Omega_p-\dot{\phi}_p$).  Conventionally, $Q$ has been assumed to be either
independent of the forcing frequency or inversely proportional to it
\citep{1968ARA&A...6..287G}.  The resulting equilibrium spin-orbit ratios as a
function of eccentricity for each of these cases are plotted in Figure
\ref{figure:spinorbit}.

Probably both of these assumptions for the behavior of $Q$ are too simple.  In
particular, if the behavior of $Q$ changes under a varying tidal forcing
frequency ($\Omega_p-\dot{\phi}_p$ is a function of time), then the tidal
equilibrium rotation rate would differ significantly from that plotted in Figure
\ref{figure:spinorbit}.  Measurement of rotational rates of eccentric extrasolar
planets in tidal equilibrium could, in principle, either differentiate between
these two models or suggest other frequency dependences, shedding light on the
yet unknown mechanism for tidal dissipation within giant planets.

\section{ROTATION AND OBLATENESS}  \label{section:rotation.oblateness}

Rotation affects the shape of a planet via two mechanisms:  gravity must
provide  centripetal acceleration, thus the higher velocity at the equator
causes the planet to bulge by transfer of mass from polar regions; and,
secondarily, the redistributed mass alters the planet's gravitational field and
attracts even more mass toward the equatorial plane.  The ratio of the required
centripetal acceleration at the equator to the gravitational acceleration, $q$,
represents the relative importance of the centripetal acceleration term:
\begin{equation} \label{eq:q} q = \frac{\Omega^2 R_{eq}^3}{G M_p} \mathrm{,}
\end{equation} where $\Omega$ is the rotation rate in radians per second, $M_p$
is the mass of the planet, and $R_{eq}$ is the planet's equatorial radius
\citep{SolarSystemDynamics}.  

We use the Darwin-Radau relation to relate rotation and oblateness accounting
for the gravitational pull of the shifted mass:
\citep{SolarSystemDynamics}:
\begin{equation} \label{eq:Darwin-Radau}
\mathbb{C} \equiv \frac{C}{M_p R_{eq}^2} = 
\frac{2}{3} 
\left[ 1 - \frac{2}{5}\left( \frac{5}{2}\frac{q}{f}-1 \right) ^{1/2} \right] 
\end{equation} 
where $C$ is the planet's moment of inertia around the rotational axis and
$\mathbb{C}$ is shorthand for $C M_p^{-1} R_{eq}^{-2}$.  The Darwin-Radau
relation is exact for uniform density bodies ($\mathbb{C}=0.4$), but is only an
approximation for gas giant planets \citep[$\mathbb{C}\sim0.25$;][]{HubbardsBook}.

By combining Equation \ref{eq:q} and Equation \ref{eq:Darwin-Radau}, we
arrive at a relation for rotation rate, $\Omega$, as a function of oblateness,
$f$:
\begin{equation} \label{eq:omega(f)}
\Omega = \sqrt{\frac{f G M_p}{R_{eq}^3} 
\left[\frac{5}{2} \left(1-\frac{3}{2}\mathbb{C}\right)^2 + \frac{2}{5}\right]}
 \mathrm{.}
\end{equation}
For our solar system, the Darwin-Radau relation yields rotation periods accurate
to within a few percent (Table \ref{table:Darwin-Radau}) using model-derived
moments of inertia from \citet{1989Icar...78..102H}.  

Extending the Darwin-Radau relation to extrasolar planets requires estimation
of the appropriate moment of inertia coefficients, $\mathbb{C}$, for those
planets.  For transiting planets whose masses and radii are known, we use a
self-consistent hydrodynamic model of the planet and assumptions about its
composition to estimate $\mathbb{C}$ following
\citet{Fortney.JupSatEvolution}.  To first order, $\mathbb{C}$ is independent
of oblateness due to similar symmetry around the rotational axis, so our
hydrodynamic model does not need to explicitly incorporate oblateness.  The
Darwin-Radau relation and spherically symmetric hydrodynamic models provide
sufficient precision for the current work; however, to estimate the oblateness
as a function of rotation more robustly, a two-dimensional interior model
involving both rotational and gravitational forces should be used
\citep[e.g.,][]{1989Icar...78..102H}.

Under the spherically symmetric assumption, we calculate the $\mathbb{C}$ of
Jupiter to be $0.277$ with no core and $0.253$ with a $20~\mathrm{M_\oplus}$
core, and we calculate the $\mathbb{C}$ of Saturn to be $0.225$ with a core
(we are unable to calculate the interior structure of Saturn without a core
due to deficiencies in our knowledge of the equation of state).  Using these
moments of inertia instead of the measured ones listed in Table
\ref{table:Darwin-Radau} yields similar rotation errors of a few percent.  We
apply our model to generic $1.0~R_{Jup}$ extrasolar giant planets of varying
masses and architectures in Figure \ref{figure:C_MR2}.

For HD209458b, our models calculate $\mathbb{C}$ to be $0.218$ with
no core and $0.185$ with a $20~\mathrm{M_\oplus}$ core.  To
estimate the oblateness of HD209458b assuming synchronous rotation, we rearrange
Equation \ref{eq:omega(f)} to solve for $f$,
\begin{equation} \label{eq:f(omega)}
f ~ = ~ \frac{\Omega^2~R^3}{G~M_p}
\left[\frac{5}{2} \left(1-\frac{3}{2}\mathbb{C}\right)^2 + \frac{2}{5}\right]^{-1}
 \mathrm{,}
\end{equation}
and then use the synchronous rotation rate $\Omega~=~2.066\times10^{-5}~
\mathrm{radians~per~second}$ to obtain $f=0.00285$ with no core and $f=0.00256$
with a $20~\mathrm{M_\oplus}$ core.  These results imply an equator-to-pole
radius difference of $\sim200~\mathrm{km}$, which, although small, is still
comparable to the atmospheric scale height at $1~\mathrm{bar}$, $\sim
700~\mathrm{km}$.

\citet{2002A&A...385..166S} suggest that zonal winds on HD209458b may operate at
speeds up to $\sim 2~\mathrm{km~s^{-1}}$ in the prograde direction, and
\citet{2002A&A...385..166S} go on to show that these winds might then spin up
the planet's interior, possibly to commensurate speeds of several hundred
$m~s^{-1}$ up to a few $km~s^{-1}$ (though the model of
\citet{2002A&A...385..166S} does not treat the outer layers and interior
self-consistently).  This speed is a non-negligible fraction of the orbital
velocity around the planet at the surface, $30~\mathrm{km~s^{-1}}$, and is also
comparable to the planet's rotational velocity at the equator,
$2.0~\mathrm{km~s^{-1}}$.  As such, if radiatively driven winds prove to be
important on HD209458b they would affect the planet's oblateness.  We use the
rotation rate implied by the \citet{2002A&A...385..166S} calculations to provide
an upper limit for the expected oblateness of HD209458b.  If the entire planet
were spinning at its synchronous rate plus $2~\mathrm{km~s^{-1}}$ at the cloud
tops, the rotational period would be halved to $1.8~\mathrm{days}$, with a
corresponding oblateness of $0.0109$ and $0.0098$ for the no core and core
models respectively.

During revision of this paper, \citet{OGLE-TR-56b.discovery.2003} announced the
discovery of a second transiting planet.  This new planet, OGLE-TR-56b, has a
radius of $1.3~\mathrm{R_{Jup}}$, a mass of $0.9~\mathrm{R_{Jup}}$, and an
extremely short orbital period of $1.2~\mathrm{days}$.  Although these
parameters are less constrained than those for HD209458b, we proceed to
calculate that the oblateness of this new object should be 0.017 with no core
and 0.016 with a $20~\mathrm{M_{\oplus}}$ core, for $\mathbb{C}$ of 0.228 and
0.204 respectively. 

Current ground-based transit searches detect low-luminosity objects like brown
dwarfs and low-mass stars in addition to planets.  However, the high surface
gravity for brown dwarfs and lower main sequence stars leads to low values of $q$
and very small oblatenesses for those objects.  For a $13\mathrm{M_{Jup}}$ brown
dwarf with $1.0\mathrm{R_{Jup}}$ in an HD209458b-like $3.52$ day orbit, the
expected oblateness is only $0.00007$.  Measuring oblateness will therefore only
be practical for transiting planets and not for other transiting low-luminosity
bodies.

We also note that the measurement of oblateness along with an independent
measurement of a planet's rotation rate $\Omega$ would determine the planet's
moment of inertia.  This would provide a direct constraint on the planet's
internal structure, possibly allowing inferences regarding the planet's bulk
helium fraction and/or the presence of a rocky core.

\section{OBLATENESS AND TRANSIT LIGHTCURVES}

\subsection{Transit Anatomy}

\citet{2001ApJ...552..699B} (hereafter BCGNB) investigated the detailed structure
of a transit lightcurve while studying the \emph{Hubble Space Telescope}
lightcurve of the HD209458b transit.  Figure \ref{figure:schematic} relates
transit events to corresponding features in the lightcurve, modeled after BCGNB
Figure 4.  The flux from the star begins to drop at the onset of transit, known
as the first contact.  As the planet's disk moves onto the star, the flux drops
further, until at second contact the entire planet disk blocks starlight.  Third
contact is the equivalent of second contact during egress, and fourth contact
marks the end of the transit.  Due to  stellar limb darkening the planet blocks
a greater fraction of the star's light at mid-transit than at the second and
third contacts, leading to curvature at the bottom of the transit lightcurve.

For a given stellar mass, $M_*$, the total transit duration, $l$, is a function
of the transit chord length and the orbital velocity.  We assume a circular
orbit, which fixes the planet's orbital velocity.  The chord length depends on
the stellar radius, $R_*$, and the impact parameter, $b$. The impact parameter
relates to $i$, the inclination of the orbital plane relative to
the plane of the sky, as
\begin{equation}
\label{eq:b}
b = \frac{\left| a \cos (i) \right|}{R_*}~\mathrm{,}
\end{equation}
where $a$ is the semimajor axis of the planet's orbit.   

The duration of ingress and egress, $w$, is the time between first and second
contact, and is a function of $R_p$, $R_*$, and $b$ (or $i$).  Transits with $b
\sim 0$ have smaller $w$ than transits with higher $b$.  For transits with $b
\sim 1$, called grazing transits, $w$ is undefined because there is no second or
third contact.

The total transit depth, $d$, fixes the ratio $R_p/R_*$, where $R_p$ is
the radius of the planet (except in the case of grazing transits). 

The magnitude of the curvature at the bottom of the transit lightcurve, $\eta$,
determines the stellar limb darkening.  We use limb darkening parameters $u_1$
and $u_2$, or $c_1$ and $c_2$, which we define mathematically in
Section \ref{section:transitmethods}.

\subsection{Methods}   \label{section:transitmethods}

We calculate theoretical transit lightcurves by comparing the amount of stellar
flux blocked by the planet to the total stellar flux.  The relative emission
intensity across the disk of the star is greatest in the center and lowest
along the edges as a result of limb darkening.  Many parameterizations of limb
darkening exist \citep[see][]{2000A&A...363.1081C}; however, we use the
one proposed by BCGNB because it is the most appropriate for planetary transits.

The emission intensity at a given point on the stellar disk, $I$, is
parameterized as a function of $\mu = \cos(\sin^{-1}(\rho/R_*))$, where $\rho$ is the
projected (apparent) distance between the center of the star and the point in
question.  BCGNB defined a set of two limb darkening coefficients, which we call
$c_1$ and $c_2$, that are are equivalent to
\begin{equation}
\label{eq:limbdarkening}
\frac{I(\mu)}{I(1)} = 1 - c_1\frac{(1-\mu)(2-\mu)}{2}+c_2\frac{(1-\mu)\mu}{2}
~\mathrm{.}
\end{equation}
The advantage of this limb darkening function is that $c_1$ describes the
magnitude of the darkening, while $c_2$ is a correction for curvature.  This
makes the BCGNB coefficients particularly useful for fitting transit
observations because a good fit to data can be achieved by fitting only for the
$c_1$ coefficient. 

Our algorithm for calculating the lightcurve takes advantage of the symmetry
inherent in the problem: that $I(\mu)$ depends only on $\rho$ and not on the
angle around the star's center, $\theta$.  We evaluate the apparent stellar flux
at time $\tau$, $F(\tau)$, relative to the out-of-transit flux $F_0$, by
subtracting the amount of stellar flux blocked by the planet from $F_0$:
\begin{equation}
\label{eq:Finf}
F_0 = \int_0^{R_*} 2\pi I(\rho)~d\rho  ~\mathrm{,}
\end{equation}
\begin{equation}
\label{eq:Fblocked}
F_{blocked} = \int_0^{R_*} 2\pi I(\rho)~x(\rho,\tau)~d\rho   ~\mathrm{,}
\end{equation}
and
\begin{equation}
\label{eq:integralalgorithm}
F(\tau) = \frac{F_0 - F_{blocked}}{F_0}~\mathrm{,}
\end{equation}
where $x(\rho,\tau)$ is the fraction of a ring of radius $\rho$ and width
$d\rho$ covered by the planet at time $\tau$.  In effect, we split the star up into
infinitesimally small rings and add up the fluxes in Equation \ref{eq:Finf},
then we determine how much of each of these rings is covered by the planet in
Equation \ref{eq:Fblocked}.  We calculate the integrals numerically using
Romberg's method \citep{NumericalRecipes}; $x(r,t)$ is evaluated numerically as
well --- there is no closed form general analytical solution to the intersection
of an ellipse and a circle.  

This algorithm is more efficient than the raster method used by
\citet{2001ApJ...560..413H} for planets treated as opaque disks because the use
of symmetry and Romberg integration minimize the number of computations of the
stellar intensity, $I(\mu)$.

 \begin{figure} 
\plotone{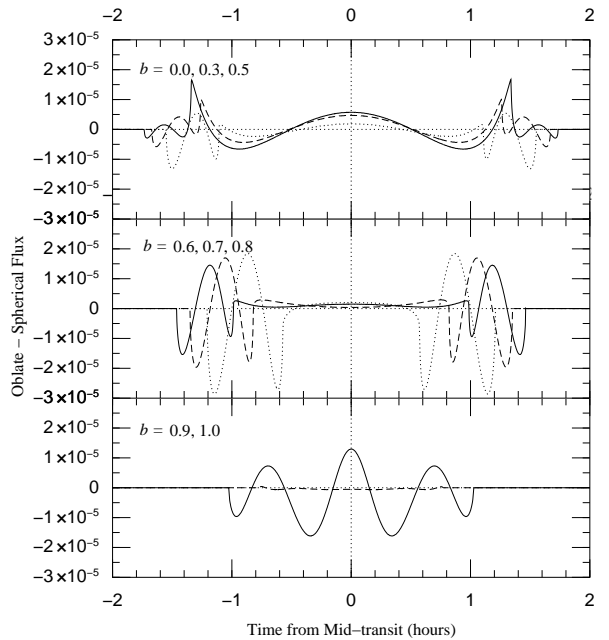}
\epsscale{0.9}
\caption{Detectable difference between the lightcurve of an oblate ($f=0.1$) planet and the
best-fit spherical model, fitting for $R_*$, $R_p$, $b$, and $c_1$.  Higher $b$
combined with larger values for $R_*$ and $R_p$ simulate the lenghened ingress
and egress of an oblate planet, diminishing the difference between oblate and
spherical planets from Figure \ref{figure:transit_effects} for planets with $b <
0.7$ (upper panel: solid line, $b = 0.0$; dashed line, $b = 0.3$; dotted
line, $b = 0.5$).  For planets near $b = 0.7$, the length of ingress and egress
cannot be simulated by higher $b$, and as a result the transit signal is highest
for these planets (middle panel: solid line, $b = 0.6$; dashed line, $b =
0.7$; dotted line, $b = 0.8$).  Above the critical value, $b > 0.7$, the oblate
planet's signal can be simulated by lowering $b$ for the spherical planet fit,
reducing the detectability of oblateness (lower panel: solid line, $b = 0.9$;
dashed line, $b = 1.0$).  It is very difficult to determine the oblateness
for planets involved in grazing transits ($b \sim 1.0$).  The magnitude of the
detectability difference is proportional to $f$ to first order, hence to
estimate the detectability of a planet with arbitrary oblateness, multiply the
differences plotted here by $ \frac{f}{0.1} \frac{R_p^2}{R_{HD209458b}^2}$.  
\label{figure:4param}}
\end{figure}

\begin{figure} 
\plotone{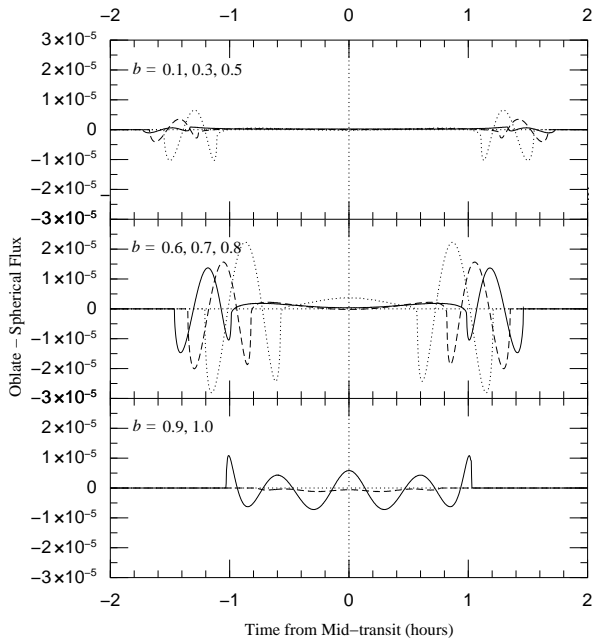}
\epsscale{0.9}
\caption{Detectable difference between the lightcurve of an oblate ($f=0.1$) planet and its
best-fit spherical model with 5 parameters: $R_*$, $R_p$, $b$, $c_1$, and $c_2$. 
As in Figure \ref{figure:4param}, differing values for $R_*$, $R_p$, and $b$ for a
spherical planet transit can allow it to mimic the transit of an oblate planet. 
With the addition of $c_2$, the fit is much better for planets transiting at
low impact parameter.  (Upper panel: solid line, $b = 0.0$; dashed line,  $b =
0.3$; dotted line, $b = 0.5$.  Middle panel:  solid line, $b = 0.6$; dashed
line, $b = 0.7$; dotted line, $b = 0.8$.  Lower panel:  solid line, $b = 0.9$;
dashed line, $b = 1.0$.)  The differences for the 5 parameter, like those for
the 4 parameter model, vary as $\frac{f}{0.1}\frac{R_p^2}{R_{HD209458b}^2}$.
\label{figure:5param}}
\end{figure}
 
\subsection{Results} \label{section:sameparams.results}

To illustrate the effect oblateness has on a transit lightcurve, we calculate
the difference between the lightcurve of an oblate, zero obliquity planet and
that of a spherical planet with the same cross-sectional area.  For
familiarity, we adopt the transit parameter values measured by BCGNB for the
HD209458b transit: $R_p=1.347\mathrm{R_{Jup}}$, $R_*=1.146\mathrm{R_\odot}$, 
and $c_1=0.640$.  Plots of the oblate-spherical differential as a function of
impact parameter are shown in Figure \ref{figure:transit_effects}.

For nearly central transits, an oblate planet encounters first contact before,
and second contact after, the equivalent spherical planet.  This situation causes
the oblate planet's transit lightcurve initially to dip below that of the
spherical planet.  However, near the time when the planet center is covering the
limb of the star, each planet blocks the same apparent stellar area and the
stellar flux difference is zero.  As the two hypothetical transits approach
second contact, this trend reverses and the spherical planet starts blocking
more light than the oblate one until the oblate planet nearly catches up at
second contact.  Between second and third contacts, the lightcurve differences
are slightly nonzero because the two planets cover areas of the star with
differing amounts of limb darkening.  The lightcurves are symmetric, such that
these effects repeat themselves in reverse upon egress.

At high impact parameters (nearly grazing planet transits) the opposite occurs. 
First contact for the oblate planet occurs after that for the spherical planet,
because the point of first contact on the planet is closer to the pole than to
the equator.  In this scenario, the oblate planet's transit flux starts higher
than, becomes equivalent to, and then drops below that of the spherical planet
before returning to near zero for the times between second and third contact. 
In the case of a grazing transit, this sequence is truncated because there is no
second or third contact.  

The boundary between these two regions occurs when the local oblate planet
radius at the point of first contact is equal to $R_{ea}$, the radius of the
equilvalent spherical planet.  For planets that are small compared with the
sizes of their stars ($R_p \ll R_*$) and that have low oblateness ($f \lesssim
0.1$), this transition occurs when $\theta = \pi/4$ (from Figure
\ref{figure:schematic}) and $b=\sqrt{2}/2=0.707$.  The flux difference between
transits of oblate planets and that for spherical planets, all else being equal,
is at a minimum at this point and deviates from zero because the rate of change
in stellar area covered is different for the two planets.  First, second, third,
and fourth contacts all occur at nearly the same time for each planet.  However,
if all else is not equal, as is usually the case since the stellar parameters
are poorly constrained, then this result is misleading and does not represent
the detectability of planetary oblatness.

\begin{figure} 
\plotone{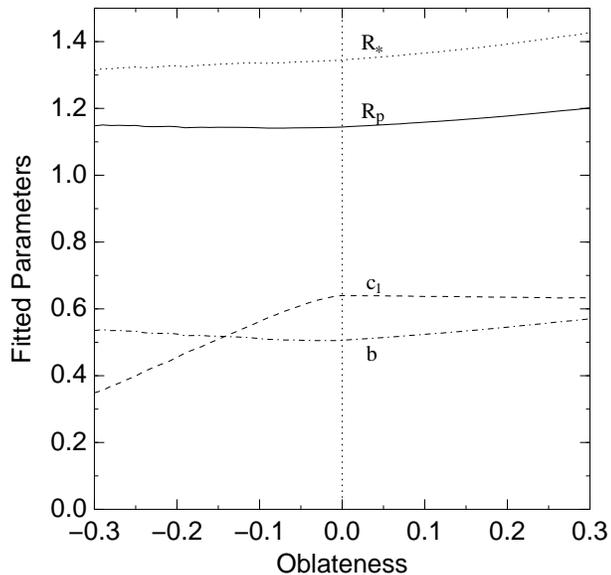}
\epsscale{0.9}
\caption{Best-fit parameters resulting from fitting a simulated transit
lightcurve for a planet with oblateness $f$ using a spherical planet transit
model.  Due to the degeneracy in measuring transit parameters $R_*$, $R_p$, and
$b$, and oblateness $f$, the best spherical fit to the oblate data have larger
radii and higher impact parameters than the actual values (see Section
\ref{section:q.eq.0}).  Objects with negative values of $f$ are prolate, a
physically unreasonable proposition that we include here for completeness.
\label{figure:p.vs.f}}
\end{figure}

\begin{figure} 
\epsscale{0.9}
\plotone{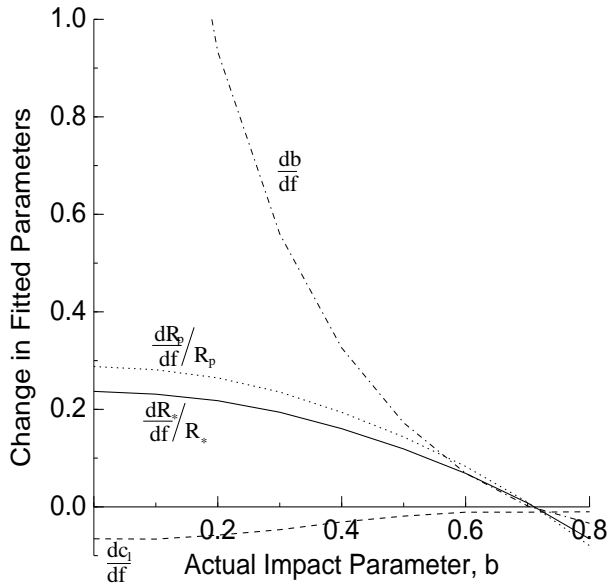}
\caption{Here we plot the change of fitted spherical model parameters with
oblateness as a function of impact parameter.  The derivitive of impact
parameter with respect to oblateness continues off the top of this graph to
$3.0$ at $b=0.0$.  An oblate planet with zero obliquity transiting near the
center of the star induces much higher deviations of the fitted parameters from
the actual parameters than would a planet transiting near the critical impact
parameter, $b=0.707$.
\label{figure:dp.df.vs.b}} 
\end{figure}

\section{TRANSIT LIGHTCURVES AND EXOPLANETS}

To test whether these flux differences are detectable, we use a
Levenberg-Marquardt curve-fitting algorithm adapted from
\citet{NumericalRecipes} to fit model transits to both the \emph{HST}
HD209458b lightcurve and hypothetical model-generated lightcurves by minimizing
$\chi^2$.  As a test, we fit the \emph{HST} HD209458b transit lightcurve and
obtain $R_*=1.142\mathrm{R_\odot}$, $R_p=1.343\mathrm{R_{Jup}}$, $i=86.72^o$
($b=0.504$), $c_1=0.647$, and $c_2=-0.065$ with a reduced $\chi^2$ of $1.05$,
consistent with the values obtained by BCGNB.

In order for planetary oblateness to have a noticeable effect on a transit
lightcurve, it must be distinguishable from the lightcurve of a spherical
planet.  For a spherical planet transit model, the combination of transit
parameters that correspond to the lightcurve that best simulates the data
from an actual oblate planet transit become the measured values, and these
measured quantities may not be similar to the actual values.  Therefore
to consider the detectability of planetary oblateness, we compare
oblate planet transit lightcurves to those of the best-fit spherical planet
lightcurve instead of to the lightcurve of a planet that differs from the
actual values only in the oblateness parameter (as we did in Section
\ref{section:sameparams.results}).

\begin{figure} 
\plotone{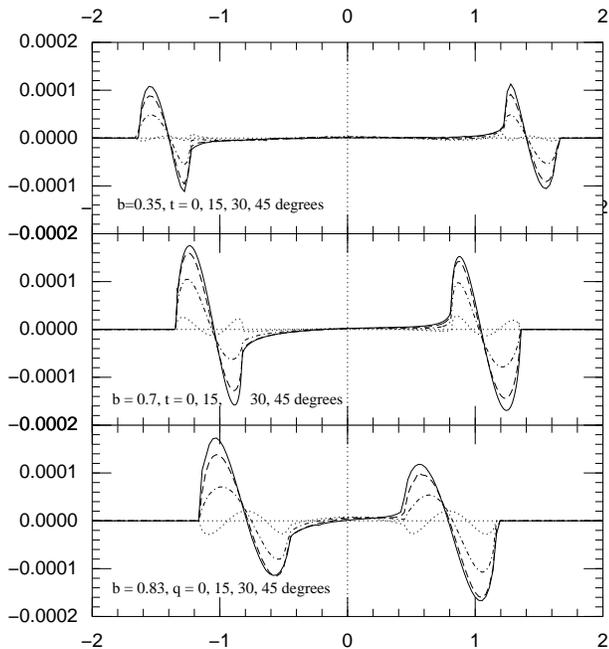}
\caption{Detectability of oblateness and obliquity relative to the best-fit
spherical model for planets with nonzero obliquity.  The top panel shows the
detectability difference for $b=0.35$ and $t=0^\circ$ (dotted line),
$t=15^\circ$ (dot-dashed line), $t=30^\circ$ (dashed line), and $t=45^\circ$
(solid line).  The middle panel represents the same varying obliquities
calculated for $b=0.7$, and the bottom panel shows the differences for
$b=0.83$.  The shape of the difference is qualitatively similar for each case. 
The difference is maximized for $t=45^\circ$ and $b=0.7$, and falls off as the
parameters near $t=0^\circ$, $t=90^\circ$, $b=0.0$, and $b=1.0$.  Due to the
symmetry of the problem, obliquities the ones shown plus $90^\circ$ are the time
inverse of the differences shown.  
\label{figure:obliq.detectability}}
\end{figure}

\subsection{Zero Obliquity} \label{section:q.eq.0}

We compare a model transit of an oblate planet ($f=0.1$) with zero obliquity,
as is the case for a tidally evolved planet, to the transit of the
\emph{best-fit} spherical planet in Figure \ref{figure:4param} and Figure
\ref{figure:5param}.  The oblate planet transit signature is muted in each when
compared to Figure \ref{figure:transit_effects} due to a degeneracy between the
fitted parameters $R_*$, $R_p$, $b$, and the oblateness $f$.  This degeneracy
is introduced by the unconstrained nature of the problem:  in essence, we are
trying to solve for 5 free parameters,  $R_*$, $R_p$, $b$, $c_1$, and $f$,
given just 4 constraints, $d$, $w$, $l$, and $\eta$ assuming (as BCGNB did)
knowledge of the stellar mass $M_*$.  Without assuming a value for $M_*$,
absolute timescales for the problem vanish, yielding a similar conundrum of
solving for $R_p/R_*$, $b$, $c_1$, and $f$ given only $d$, $\eta$, and $w/l$.
The previous two sentences are intended only to be simplifications, as at
higher photometric precision more information about the conditions of the
transit is available.  Hereafter we assume knowledge of $M_*$, though this
analysis could also have been done without this assumption, or with an assumed
relation between $M_*$ and $R_*$ as proposed by \citet{2002ApJ...569..451C}.  
Changes in $R_*$, $R_p$, and $b$ mimic the signal of an oblate planet by
altering the ingress and egress times while maintaining the total transit
duration by keeping the chord length constant.

For planets transiting at low impact parameter ($b < 0.7$), an oblate planet's
longer ingress and egress (higher $w$ from Figure \ref{figure:schematic}) are
fit better using a spherical model with a higher impact parameter than actual,
thus lengthening the time between first and second contact.  Since for a given
star transits at higher impact parameter have shorter overall duration, the
best-fit spherical model has a larger $R_*$ than the model used to generate the
data to maintain the chord length, and thus a larger $R_p$ to maintain the
overall transit depth.

Similarly, for simulated lightcurve data from a transiting oblate planet at high
impact parameter ($b > 0.7$), the best-fit spherical model has a lower impact
parameter than the simulated planet to increase the duration of ingress and
egress.  For these planets, the fitted spherical parameters have smaller $R_*$
and $R_p$ than the simulated planet to maintain the character of the rest of the
lightcurve.

Oblate planets that have impact parameters near the critical value, $b \sim
0.7$, cannot be as easily fit using a spherical model because changes in the
impact parameter of the fit cannot increase the duration of ingress and egress. 
For these planets, the difference between the oblate planet transit lightcurve
and the best-fit spherical planet transit lightcurve is maximized, providing the
largest possible photometric signal with which to measure oblateness.

At present, it is necessary to fit for $R_*$ and stellar limb-darkening
parameters because of our limited knowledge of these values for the host stars
of transiting planets.  If $R_*$ were known to sufficient accuracy (less than
1\%), then with knowledge of $M_*$ the degeneracy between $R_*$, $R_p$, and $b$
would be broken, allowing measurement of planetary oblateness without fitting. 
However, without assumptions about the stellar mass, knowledge of $R_*$ would
only help to constrain $M_*$.  \citet{2002ApJ...569..451C} show that, for the
star HD209458, current evolutionary models combined with transit data serve to
measure the stellar radius to a precision of only 10\%.

In Figure \ref{figure:4param} we plot the difference between the transit
lightcurve of a hypothetical planet with the characteristics of HD209458b but an
oblateness $f=0.1$, and that planet's best-fit spherical model fitting for 
$R_*$, $R_p$, $b$, and $c_1$.  For low values of the impact parameter, $b \le
0.5$, the best-fit spherical lightcurve emulates the oblate planet's ingress and
egress while leaving a subtly different transit bottom due to the planet
traversing a differently limb-darkened chord across the star.  The magnitude of
the difference is approximately a factor of $10$ smaller than the non-fit
difference from Figure \ref{figure:transit_effects}.  Near the critical impact
parameter, $b=0.7$, the transit lightcurve bottoms are very similar since the
best-fit $b$ is very similar to the actual oblate planet's impact parameter;
however, the ingress and egress differ in flux by a few parts in $10^5$ for
$f=0.1$.  For grazing transits, $b\sim1.0$, the best-fit spherical model's
lightcurve is indistinguishable from the oblate planet transit lightcurve even
at the $10^{-6}$ relative accuracy level.  

Figure \ref{figure:5param} shows the same differences as in Figure
\ref{figure:4param}, but with a second stellar limb darkening parameter, BCGNB's
$c_2$, also left as a free parameter in the fit.  Including $c_2$ in the fit
allows the fitting algorithm to match the transit bottom (the time between second
and third contacts) better.  This effect leads to excellent fits of oblate planet
transits using spherical planet models and reduces the detectable difference to
less than one part in $10^5$ for $0.0 < b < 0.5$ and $b > 0.9$ using HD209458
stellar and planetary radii and an oblateness of $f=0.1$.  For $b=0.0$ and
$b=1.0$ the spherical model emulates an oblate transit particularly well, with
the differences between the two being only a few millionths of the
stellar flux.

In both the 4- and 5-parameter zero-obliquity models, it is easiest to measure
the effects of oblateness on the transit light curve for transits near the
critical impact parameter.  For HD209458 values with $f=0.1$, the oblateness
signal then approaches $3\times10^{-5}$ for tens of minutes during ingress and
egress and peaks near $b=0.8$ instead of the critical value of $b=0.7$ due to
the finite radius of HD209458b relative to its parent star.

Based on observations of the Sun, \citet{1997pbss.conf..153B} expect the
intrinsic stellar photometric variability on transit timescales to be $\sim
10^{-5}$.  \citet{2002ApJ...575..493J} used 5 years of 3 minute resolution SOHO
spacecraft data to deduce the noise power spectrum of the Sun.   These noise
effects are close enough in magnitude to the transit signal so as to affect the
detectability of a transiting oblate planet.  

Future high-precision space-based photometry missions such as MOST and Kepler
may be able to detect the effect for highly oblate transiting extrasolar
planets, but the S/N ratio could be so low as to make unambiguous measurements of
oblateness diffucult to obtain.

If an observer were to fit photometric timeseries data from a transit of an
oblate planet without fitting explicitly for the oblateness $f$ (as, for
instance, if the precision is insufficient to measure $f$), the planet's
oblateness will be manifest as an astrophysical source of systematic error in the
determination of the other transit parameters.  Figure \ref{figure:p.vs.f} shows
the effect that oblateness has on the stellar radius, planetary radius, impact
parameter, and limb darkening coefficient for HD209458b.  This variation is a
strong function of the planet's impact parameter.  In Figure 
\ref{figure:dp.df.vs.b}, we show the how this systematic variation changes as a
function of the initial impact parameter for the HD209458b system.  As a severe
but still physical example, an HD209458b-like planet with oblateness $f=0.1$ that
transits at $b=0.0$ would be measured to have radii $2\%$ above actual and an
impact parameter nearly $0.3$ above the real impact parameter.

\subsubsection{HD209458b}

To illustrate the robustness of the degeneracy between $R_*$, $R_p$, $b$, and $f$
discussed in Section \ref{section:q.eq.0}, we fit the \emph{HST} lightcurve from
BCGNB using a planet with a fixed, large oblateness of $0.3$(!).  The  best-fit
parameters were $R_*=1.08~\mathrm{R_\odot}$, $R_p=1.26~R_{Jup}$, $b=0.39$, and
$c_1=0.633$ with a reduced $\chi^2=1.06$ --- indistinguishable in significance
from the spherical planet model!  Although unlikely, a high actual oblateness for
HD209458b could alter the measured value for the planet's radius into better
agreement with theoretical models.

The expected detectability of oblateness for HD209458b is extremely low. 
During ingress and egress the transit lightcurve for HD209458b should differ
from that of the best-fit spherical model by only one part in $10^6$ for $f=0.01$
and at the level of $3\times10^{-7}$ if $f=0.003$.  For comparison, the BCGNB
\emph{HST} photometric precision is $1\times10^{-4}$.

Although the systematic error in the determination of transit parameters can be
important for highly oblate planets, it is not at all significant for
HD209458b.  Assuming HD209458b has an oblateness of $f=0.003$ as calculated in
Section \ref{section:rotation.oblateness}, the fitted radii are only $\sim
0.05\%$ above the actual radii, and the fitted impact parameter is $0.0006$
above what the actual impact parameter should be.

\subsubsection{OGLE-TR-56b}

Measuring the oblateness of the new transiting planet OGLE-TR-56b
\citep{OGLE-TR-56b.discovery.2003}, $f=0.016$ (see
Section \ref{section:rotation.oblateness}), would require photometric precision
down to at least $4\times10^{-6}$.  Since the impact parameter for this
object is as yet unconstrained, the above precision corresponds to $b=0.7$. 
For other transit geometries, higher precision photometry would be necessary to
measure the oblateness of this object.

\subsection{Nonzero Obliquity}

We plot the detectability of oblateness and obliquity for planets with nonzero
obliquity in Figure \ref{figure:obliq.detectability}.  Here we define the
projected obliquity (which we refer to as just obliquity hereafter), or axial
tilt $t$, as the angle between the orbit angular momentum vector and the
rotational angular momentum vector projected into the sky plane, measured
clockwise from the angular momentum vector (see Figure \ref{figure:schematic}). 
The major effect of nonzero obliquity is to introduce an asymmetry into the
transit lightcurve \citep{Hui.Seager.2002}.

The plots in Figure \ref{figure:obliq.detectability} are difference plots
for planets with different obliquities and impact parameters, yet all have
the same shape qualitatively. 
In trying to fit the asymmetric ingress and egress lightcurves, the best-fit
spherical planet splits the difference between them.  For planets with $0 < t <
\pi/2$, this process yields a difference curve that initially increases as the
spherical planet covers the star at a faster rate than does the oblate
planet.  Because of the asymmetric nature of the problem, however, the egress of
the oblate planet takes longer than the spherical one, causing an initial upturn
near third contact.  The transit lightcurve difference for planets with $0 >
t > -\pi/2$ is equal to the one for $0 < t < \pi/2$, but reversed in time.

This general shape is the same for the asymmetric component of each transit
lightcurve, and it is superimposed on top of the symmetric component studied in
Section \ref{section:q.eq.0}.  The asymmetric component is maximized near the
critical impact parameter ($b=0.7$), because the planet crosses the stellar limb
with its projected velocity vector at an angle of $\pi/4$ with respect to the
limb.  For transits across the middle of the star, $b=0$, the asymmetric
component of the lightcurve vanishes as a new symmetry around the planet's path
is introduced, and the local angle between the velocity vector and the limb is
$\pi/2$.  Similarly the asymmetric planet signal formally goes to zero for
grazing transits at $b~=~1.0$, but in practice the asymmetric component is still
high for Jupiter-sized planets.

Planets with obliquities of zero ($t=0$) are symmetric and have no asymmetric
lightcurve component (see Section \ref{section:q.eq.0}).  Likewise, planets with
$t=\pi/2$ are also symmetric.  The asymmetric component is maximized for
planets with $t=\pi/4$.

Transiting planets with nonzero obliquity can break the degeneracy between
transit parameters discussed in Section \ref{section:q.eq.0}.  However, when
the obliquity is nonzero a degeneracy between projected obliquity and
oblateness is introduced:  a measured asymmetric lightcurve component of a
given magnitude could be due to a planet with low oblateness but near the
maximum detectability obliquity of $t=\pi/4$, or it could be the result of a
more highly oblate planet with a lower obliquity.  In this case only a lower
limit to the oblateness can be determined based on the oblateness for an
assumed obliquity of $\pi/4$.  This degeneracy can be broken with measurements
precise enough to determine the symmetric lightcurve component.  In addition,
transit photometry is only able to measure the projected oblateness and
obliquity of such objects due to the unknown component of obliquity along the
line-of-sight \citep{Hui.Seager.2002}, therefore the true oblateness is never
smaller than the measured, projected oblateness.

The asymmetric transit signal of an oblate planet with nonzero obliquity could
also be muddled by the presence of other bodies orbiting the planet.  Orbiting
satellites or rings could both introduce asymmetries into the transit lightcurve
that may not be easily distinguishable from the asymmetry resulting from the
oblate planet.  Satellites around tidally evolved planets are not stable
\citep{ExtrasolarMoons}, and rings around these objects may prove to be
difficult to sustain as well.  However, objects that are not tidally evolved,
those farther away from their parent stars, may potentially retain such
adornments, and their effects on a transit lightcurve could be difficult to
differentiate from oblateness.

Orbital eccentricity can also cause a transit lightcurve to be asymmetric. 
Although the eccentricity of HD209458b is zero to within measurement uncertainty
based on radial velocity measurements, future planets discovered solely by their
transits may not have constrained orbital parameters.  For these objects with
unknown eccentricity, if the eccentricity is very high then under some conditions
the velocity change between ingress and egress may be high enough to emulate the
oblateness asymmetry discussed in this section.  However, we do not explicitly
treat that situation in this paper.

\section{CONCLUSIONS}

Examining a transiting planet's precise lightcurve can allow the measurement of
the planet's oblateness and, therefore, rotation rate, beginning the process of
characterizing extrasolar planets.  To a reasonably close approximation (a few
percent), the rotation rate of an extrasolar giant planets is related to the
planet's oblateness by the Darwin-Radau relation.  Measurements of a planet's
rotation rate could constrain the tidal dissipation factor $Q$ for those
planets, as well as possibly shed light on the tidal dissipation mechanism
within giant planets based on the spin : orbit ratios of tidally evolved eccentric
planets.

The detectable effect of oblateness on the lightcurve of a planet with zero
obliquity is at best on the level of a few parts in $10^5$.  This effect may be
discernable from space with ultra high precision photometry, but could prove to
be indistinguishable from stellar noise or other confounding effects.  Accurate
independent measurements of stellar radius can break the degeneracy between the
stellar radius, planetary radius, and impact parameter, allowing for much easier
measurement of oblateness.  Without such measurements, the primary effect of
oblateness and zero obliquity when studying transit lightcurves will be to
provide an astrophysical source of systematic error in the measurement of
transit parameters.  

Planets with nonzero obliquity have higher detectabilities (up to $\sim
10^{-4}$) than planets with no obliquity, but yield nonunique determinations of
obliquity and oblateness.  In order to obtain unique obliquities and
oblatenesses for these objects, precision comparable to that needed for the zero
obliquity case is needed.

The detectability of oblateness for transiting planets is maximized for impact
parameters near the critical impact parameter of $b=0.7$.  Many transit
searches are currently underway, yielding the potential for the discovery of
many transiting planets in coming years.  Given the opportunity to
attempt to measure oblateness, our analysis suggests that the optimal
observational target selection strategy would be to observe planets around
bright stars that transit near an impact parameter of $0.7$.

\acknowledgements

The authors wish to acknowledge Bill Hubbard and Fred Ciesla for useful
conversations; Wayne Barnes, Robert H. Brown, Christian Schaller, and Paul Withers
for manuscript suggestions; and Robert H. Brown for valuble advice.

\newpage

\bibliographystyle{apj}

\newpage

\newpage 
\begin{deluxetable}{l|c|l|c|r|c}
\tablecaption{Darwin-Radau in the Solar System\label{table:Darwin-Radau}}
\tablewidth{0pt}
\tablehead{
\colhead{Planet} &
\colhead{f} & 
\colhead{$\mathbb{C}$} &
\colhead{Calculated Period} & 
\colhead{Actual Period} &
\colhead{Error}
}
\startdata
Jupiter  & 0.06487  & 0.26401\tablenotemark{1} & 10.1609 hr   & 9.92425 hr & 2.38\% \\
Saturn   & 0.09796  & 0.22037\tablenotemark{1} & 10.8817 hr   & 10.6562 hr & 2.12\% \\
Uranus   & 0.02293  & 0.2268 \tablenotemark{1} & 16.6459 hr   & 17.24 hr   & 3.45\% \\
Neptune  & 0.01708  & 0.23   \tablenotemark{*} & 16.8656 hr   & 16.11 hr   & 4.69\%  \\
Earth    & 0.00335  & 0.3308   & 23.8808 hr   & 23.9342 hr & 0.223\% \\
\enddata
\tablenotetext{1}{\citet{1989Icar...78..102H}}
\tablenotetext{*}{Assumed}
\tablecomments{
Comparison of calculated rotation rates from the Darwin-Radau approximation
(Equation \ref{eq:Darwin-Radau}) to actual rotation rates for selected planets in our solar system.}
\end{deluxetable}

\end{document}